# Persistent or repeated surface habitability on Mars during the Late Hesperian - Amazonian

Edwin S. Kite[1,*], Jonathan Sneed[1], David P. Mayer[1], Sharon A. Wilson[2]

1. University of Chicago, Chicago, Illinois, USA (* kite@uchicago.edu)
2. Smithsonian Institution, Washington, District of Columbia, USA

**Key Points:**
• Alluvial fans on Mars did not form during a short-lived climate anomaly
• We use the distribution of observed craters embedded in alluvial fans to place a lower limit on fan formation time of >20 Ma
• The lower limit on fan formation time increases to >(100-300) Ma when model estimates of craters completely buried in the fans are included

**Abstract.**
Large alluvial fan deposits on Mars record relatively recent habitable surface conditions (≲3.5 Ga, Late Hesperian – Amazonian). We find net sedimentation rate <(4-8) μm/yr in the alluvial-fan deposits, using the frequency of craters that are interbedded with alluvial-fan deposits as a fluvial-process chronometer. Considering only the observed interbedded craters sets a lower bound of >20 Myr on the total time interval spanned by alluvial-fan aggradation, >$10^3$-fold longer than previous lower limits. A more realistic approach that corrects for craters fully entombed in the fan deposits raises the lower bound to >(100-300) Myr. Several factors not included in our calculations would further increase the lower bound. The lower bound rules out fan-formation by a brief climate anomaly. Therefore, during the Late Hesperian – Amazonian on Mars, persistent or repeated processes permitted habitable surface conditions.

## 1. Introduction.

Large alluvial fans on Mars record one or more river-supporting climates on ≲3.5 Ga Mars (e.g. Moore & Howard 2005, Grant & Wilson 2012, Kite et al. 2015, Williams et al. 2013) (Fig. 1). This climate permitted precipitation-sourced runoff production of >0.1 mm/hr that fed rivers with discharge up to $10^2$ $m^3$/s (Dietrich et al. 2017, Morgan et al. 2014). Such large water discharges probably require a water activity that would permit life. The large (>10 $km^2$) alluvial fans with ≲2° slopes correspond to a relatively recent (Grant & Wilson 2011) epoch of Mars surface habitability (Williams et al. 2013). (Although small (<10 $km^2$) alluvial fans with >10° slopes formed <5 Mya on Mars (e.g. Williams & Malin 2008), in this paper we focus on the large alluvial fans). Did these large fans result from a single anomalous burst of wet conditions, such as might result from an impact or volcanic eruption? Or, do the fans record persistent or repeated wet conditions, for example as the result of a sustained warmer climate regime? Better constraints on the time span of alluvial fan formation would constrain models of Late Hesperian–Amazonian climate.

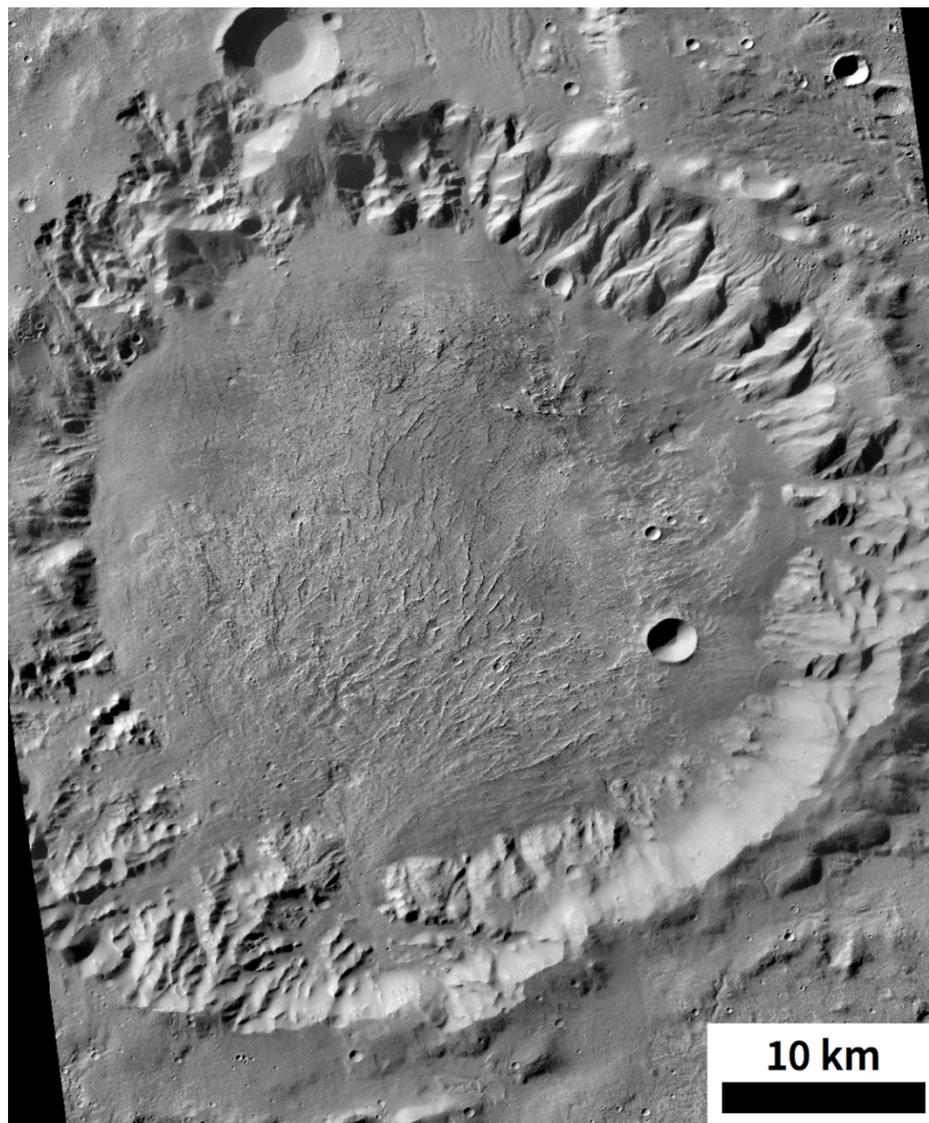

**Figure 1.** Large alluvial fan deposit on Mars (28°S 333°E). The ≲3.5 Ga age of the fans is shown by their relatively low crater density. Aeolian erosion exposes layers and channels within the deposit.

Previous work on the duration of the interval of fan build-up used sedimentology to estimate the time over which sediment transport occurred (e.g., Armitage et al. 2011, Williams et al. 2011, Palucis et al. 2014). These sedimentologic methods require assumptions about flow intermittency or sediment:water ratio, which (for almost all Mars alluvial fans) are poorly constrained (Dietrich et al. 2017). Therefore, the lower limits obtained from sedimentological methods are short – for example, ~3600 years (Morgan et al. 2014). Moreover, brief (1-100 yr) aggradation intervals have been proposed for deltas on Mars that are similar to the alluvial fans in age and volume (Kleinhans et al. 2010, Mangold et al. 2012a, Hauber et al. 2013). Another approach to estimating the interval of alluvial-fan build-up is to measure the density of craters superimposed on different fans, and use the spread of crater-retention ages for the fan surfaces as a proxy for the range of fan-formation ages. This method is not reliable, because crater counts for areas <$10^3$ km$^2$ cannot distinguish between ages of (for example) 2.5 Ga and 3.0 Ga (Warner et al. 2015). Therefore, the time span of habitable climates in the Late Hesperian-Amazonian remains an open question.

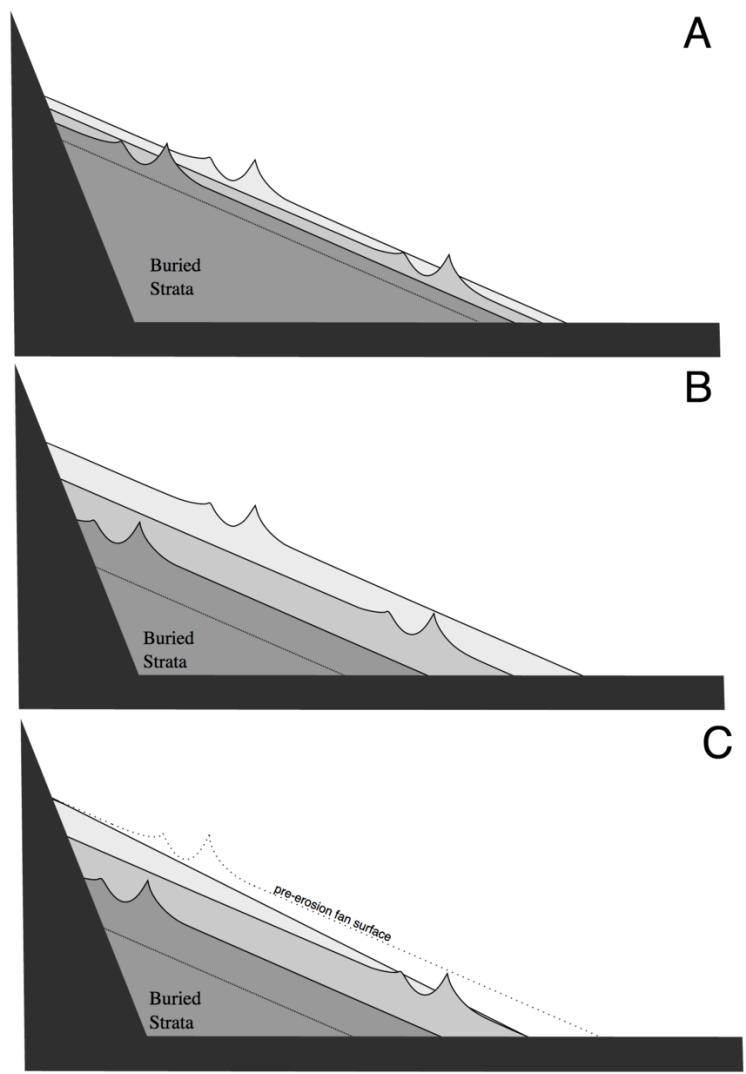

**Figure 2.** Idealized cross-section through an alluvial fan deposit. Crater density within alluvial fans may be estimated using the frequency of visible interbedded craters at the exposed surface. Case A: If fluvial sediment accumulation was modest, some craters that formed late in the history of alluvial-fan aggradation are only partially buried and are visible today (though outnumbered by postfluvial craters; not shown). Case B: If fluvial accumulation was rapid, the temporal window for partial crater burial was correspondingly brief, and few interbedded craters are visible at the exposed surface. Case C: If postfluvial erosion was severe, the areal density of exposed embedded craters of a given diameter is still proportional to the volumetric density of those craters. In either erosional or well-preserved fan scenarios, assuming steady aggradation, a count of interbedded craters constrains the fan aggradation rate.

To get a more accurate estimate of the interval over which alluvial fans formed, we used the embedded-crater method (Hartmann 1974, Kite et al. 2013a). For post-Noachian deposits on Mars, this method works as follows. Crater density on a stable planetary surface (a surface that is neither eroding nor aggrading) is proportional to exposure duration. This method may be extended to three dimensions: the total number of craters interbedded within a sedimentary deposit is proportional to the time spanned by active sedimentation (including any hiatuses) (Fig. 2). The greater the volumetric density of craters, the longer the time span of deposition. Of course, many or most of the interbedded craters may be completely buried (Fig. 2); therefore a complete count of interbedded craters is usually impossible. However, if an impact occurs near the end of active sedimentation, then the corresponding crater may be only partially buried. Smaller craters are more readily buried, and larger craters require more sediment to be completely obscured. The time needed to accumulate the population of visibly embedded (synfluvial) craters is

$$\tau_{raw,D} = N_D/(f_D a) \qquad\qquad [1]$$

where $\tau_{raw,D}$ is the minimum time required to build up the observed population of embedded craters (with minimum diameter $D$), $N_D$ is the number of observed embedded craters, $f_D$ is the past crater flux (#/km$^2$/yr), and $a$ is the count area (km$^2$). Prefluvial craters (which are overlain by fan deposits, but that formed before the start of fluvial deposition; Irwin et al. 2015) are excluded. This procedure gives a strict lower limit on the interval of fan formation; it does not account for craters within the deposit that are fully entombed. To correct for fully entombed craters, we can assume steady aggradation and divide fan thickness $Z$ by best-fit aggradation rate $W$ to get duration of fan formation $\tau_{steady}$:

$$W_D \approx 1.33 D\varphi/\tau_D \qquad\qquad [2]$$

$$\tau_{steady,D} = Z/W_D \qquad\qquad [3]$$

The numerator in [2] corresponds to the required burial depth for obliteration. The amount of burial that is required is constrained by the geometry of small impact craters (Melosh 1989, Watters et al. 2015). $\varphi$ is the obliteration depth fraction for a given crater, expressed relative to crater diameter. (In this paper, we define a crater as "obliterated" if it can no longer be identified in a high-resolution optical image; Kite & Mayer 2017). The factor of 1.33 corrects for the fact that, for any minimum-diameter $D$, the median diameter in a count will exceed $D$ – thus, $D\varphi$ is an underestimate of the required burial depth. This correction depends on the crater production function used. The correction is relatively small (1.3×–1.5×) in our size range of interest, because crater frequency falls off steeply with increasing diameter, and we represent it by a fixed factor. Post-depositional erosion of fan deposits does not affect the validity of [2-3]. A randomly oriented cut through the crater-containing volume intersects each crater with a probability proportional to that crater's size; the sample of craters partially exhumed at an erosional surface is biased towards larger impacts. Just as with partial burial, the number of craters that are exposed is proportional to the volumetric crater density (and inversely proportional to aggradation rate). Thus, the volumetric density of interbedded craters may be estimated from surface counts both on pristine fan surfaces and for fans that are deeply eroded (Kite et al. 2013a). Therefore, embedded-crater counts can be used as a Mars fluvial process speedometer.

## 2. Methods and results.

In order to set a lower bound on the time span of alluvial-fan formation, we searched $1.7 \times 10^4$ km$^2$ of previously-catalogued fans (corresponding to most of the surface area of large alluvial fans on Mars; Wilson et al. 2012) using 6m-per-pixel CTX images to scout for candidate embedded craters. Candidate craters show possible evidence of interbedding with paleochannels or other fan deposits. Each candidate feature was reviewed by three of the authors (E.S.K, D.P.M., and J.S.) for final classification. Where available, 25cm-per-pixel HiRISE images and anaglyphs, plus CTX Digital Terrain Models (DTMs) were used to inspect candidates flagged in the initial CTX survey. Each candidate feature was categorized as quality level 1, 2, 3 (representing decreasing confidence that the crater was embedded), or it was discarded. A total of 25 embedded craters were found at <30° latitude ($D = 0.08$-5.0 km; Fig. 3, Figs. S1-S2). The latitude cut was selected to avoid glaciated craters. Those embedded craters were then classified (usually with the aid of CTX DTMs) as "synfluvial," "uncertain", or "prefluvial." These craters constitute a small fraction of the total number of craters on the surfaces of the fans, most of which appear to be postfluvial (Grant & Wilson 2012).

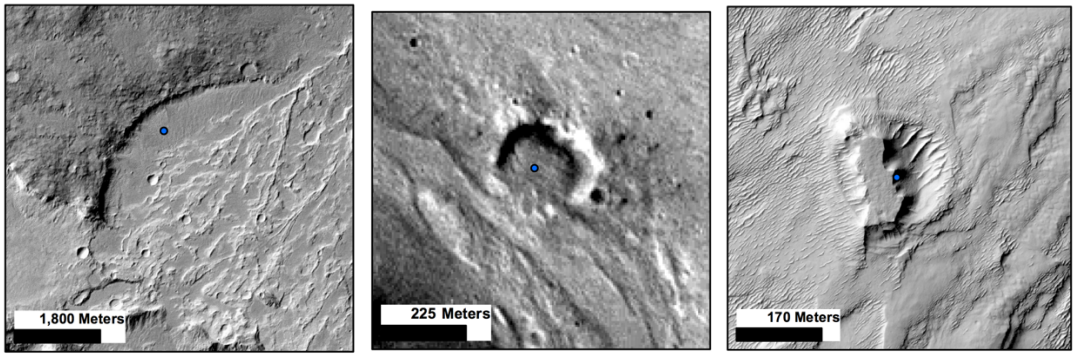

**Figure 3.** Embedded synfluvial craters within alluvial fans on Mars from our catalog (Fig. S2).

A supplementary HiRISE-only survey (570 km$^2$) was carried out to check for resolution effects. 13 embedded craters were found ($D$=0.05–0.22 km; Figure S3). For 0.08 km < $D$ < 0.16 km, HiRISE embedded-crater densities $N_D/a$ are the same (within Poisson error) as CTX embedded-crater densities. Therefore, the HiRISE check provides no evidence that our conclusions would be changed by a HiRISE re-survey of the full area covered by our CTX survey.

Diameter measurement error was estimated by blindly remeasuring craters and found to be negligible. Crater diameter change during degradation is ignored.

The contribution of false positives to our catalog is likely small. Although polygonal faulting in Earth marine sediments can produce crater-like concentric layering (Tewksebury et al. 2014), this is unlikely for Mars alluvial-fan deposits. For example, embedded craters are isolated (not space-filling), and frequently show preserved rims. However, there are certainly false negatives in our survey area: re-survey of a crater of interest found several additional candidates with scores ≤3 on panel inspection. It is possible that many embedded craters are easily identifiable as impact craters in CTX imagery, but have an expression that is indistinguishable from postfluvial craters at CTX scale. Therefore, our embedded crater densities are lower limits.

We estimated fan thicknesses by differencing CTX DTM profiles across fans and analogous profiles across parts of the same fan-hosting craters ($n$=17) that lacked fans (Palucis et al. 2014) (Table 1). We found maximum fan thickness of 1.1km, with thicknesses ~1 km common.

| Crater | Name | Estimated maximum fan thickness | Comments |
|---|---|---:|---|
| 46W 0N | Orson Welles | **900 m** | Flat floor approximation. |
| 45W 24S | Degana | 200 m | |
| 37W 18S | Luba | >200 m | Conservative estimate |
| 34W 26S | Holden | **1.1 km** | Flat floor approximation. |
| 28W 26S | Ostrov | **1 km** | Flat floor approximation. |
| 27W 28S | "SE of Ostrov" | >300 m | Conservative estimate |
| 15W 25S | - | 150 m | |
| 67E, 22S | Harris | 300 m | |
| 73E 22S | Saheki | **>1.1 km** | Flat floor approximation. |
| 74E 23S | "SE of Saheki" | 500 m | Flat floor approximation. |
| 76E 26S | Runanga | 250 m | Alternative control gives max. thickness 450 m. |
| 84E 29S | - | **900 m** | |
| 84E 33S | Majuro | 300 m – 400 m | Taken from Mangold et al. 2012b |
| 134E 1N | - | >500 m | |
| 142E 12N | Eddie | 250 m | |
| 170E 19N | Kotka | 300 m | |
| 180E 5S | - | 200 m | |

**Table 1.** Alluvial fan thickness estimates (crater-by-crater maximum estimated thickness). Thicknesses ~1 km are shown in bold.

### 3. Analysis.

The usual procedure for estimating crater-counting error is to use Poisson statistics (Michael et al. 2016). The results of this procedure are shown by the blue lines and blue error bars in Fig. 4. To generate these results, we assumed a fixed crater flux (Michael 2013), no change in atmospheric screening from today's Mars, a strong-rock target strength, and a fixed obliteration depth fraction $\varphi$ = 0.1.

The true error is larger than this because of uncertainty in:- true crater flux (Hartmann 2005, Johnson et al. 2016); target strength (Okubo 2007, Grindrod et al. 2010); filtering by a potentially thicker past atmosphere (Kite et al. 2014); the time of formation of the fans; and the amount of burial or erosion (expressed as a fraction of the crater's diameter) that is needed to prevent the crater from being detected at CTX resolution.

Therefore, we adopted conservative prior probabilities on these parameters in a Monte Carlo simulation of our lower bound that also includes Poisson error (details are given in the Supplementary Information). Specifically, we assumed (1) a factor-of-4 uncertainty in crater flux (log-uniform uncertainty between 0.5× and 2× the fluxes of Michael 2013); (2) log-uniform uncertainty in target strength between limits of 65 kPa and 10 MPa (Dundas et al. 2010); (3) log-uniform uncertainty in paleo-atmospheric pressure between limits of 6 mbar and 1000 mbar; (4) a uniform uncertainty between fan formation 2.0 Ga (low-end fan crater retention age) and 3.6 Ga (ages of large craters that host fans); and (5) a log-uniform prior for obliteration depth fraction (expressed as a fraction of $D$) from 0.05 (rim burial; Melosh 1989) and 0.2 (original crater depth; Watters et al. 2015). For each Monte Carlo trial, the effect of Poisson error is calculated analytically. Given the observations and the randomly-sampled parameters, each Monte Carlo trial yields an analytic probability for each candidate age (or each candidate aggradation rate) in each size bin. These probabilities are summed over $10^3$ Monte Carlo trials and normalized. Results are shown by the gray bands and black stars in Fig. 4.

The best-fit lower limit on the time-span of fan aggradation (Fig. 4a) increases with increasing diameter, as expected. Bins ≥1.4 km contain only 1 crater, and the Poisson uncertainty constitutes most of the total uncertainty in the lower limit. For the smaller diameters, the counting-statistics error is small compared to the total uncertainty in the lower limit, but systematic undercounting of embedded craters is most likely for craters that are smaller (and thus more easily buried and modified). For example, for 10 m diameter craters, the nominal time span to build up the observed craters is ~1 Myr, but the amount of burial needed to obliterate the crater is only ~1 m, and given fan thicknesses ~1 km (Table 1), it is very unlikely that there are zero craters entombed within the lower 99.9% of the fans. The largest >1 km-diameter bin contains 2 embedded craters and yields a 95% lower limit of >17 Myr, which we round to >20 Myr. Using the single ~5km crater found in our survey gives a >54 Myr lower bound.

Turning to the rate plot (Fig. 4b), constant aggradation rates of <(4-8) μm/year match our data. Different crater diameters probe different depth ranges and thus aggradation rate over different timescales, but our data do not require a change in aggradation rate with timescale (Jerolmack & Sadler 2007). Dividing typical fan thicknesses by this rate gives 125-250 Myr (100-300 Myr to 1 significant figure). For both plots, the Monte Carlo procedure gives limits that are more uncertain, and slightly more permissive (Fig. 4).

The time span of liquid water estimated from steady aggradation (>(100-300) Myr; Fig. 4b) is a more realistic estimate of the true fan-forming interval than the time span estimated from observed synfluvial craters (>20 Myr; Fig. 4a). For example, if the observed embedded craters represent the total embedded-crater population, then fan aggradation must have started very fast and decreased sharply near the end of fan build-up. Such an accumulation history would favor small-crater preservation relative to large-crater preservation – opposite what is observed. Furthermore, after fluvial deposition stopped, many of the fan deposits underwent aeolian erosion (e.g. Fig. 1). Because postfluvial erosion would destroy some embedded craters, the observed embedded craters are very unlikely to represent the total population of craters that formed embedded within the fans.

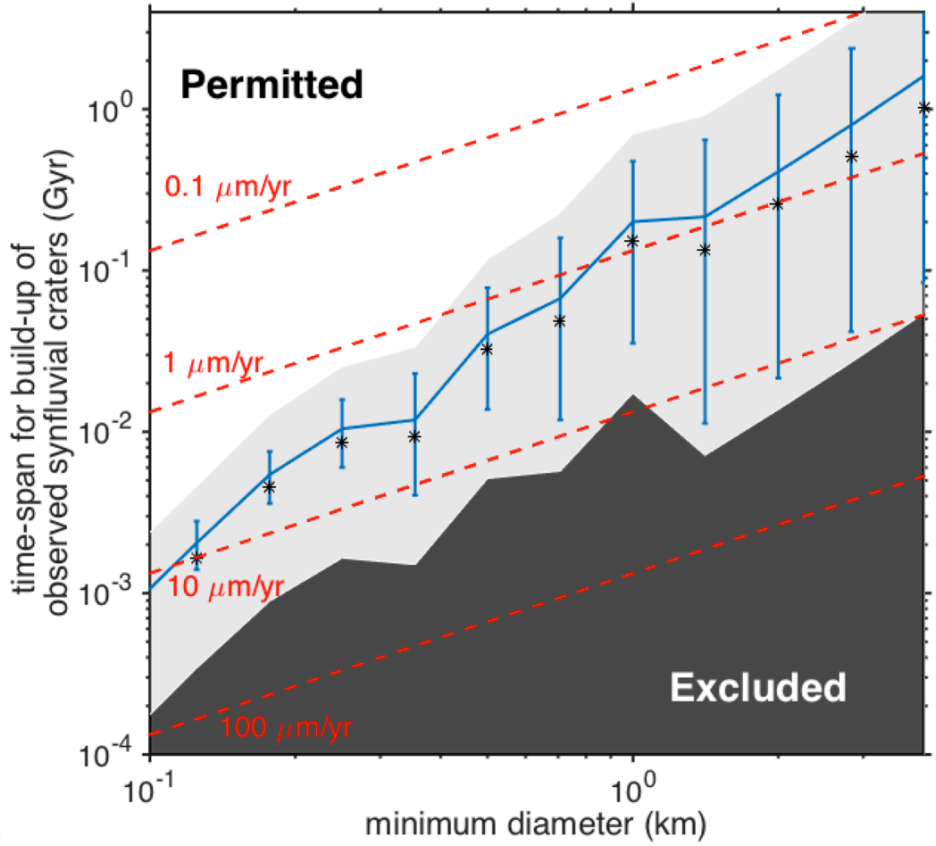

(a)

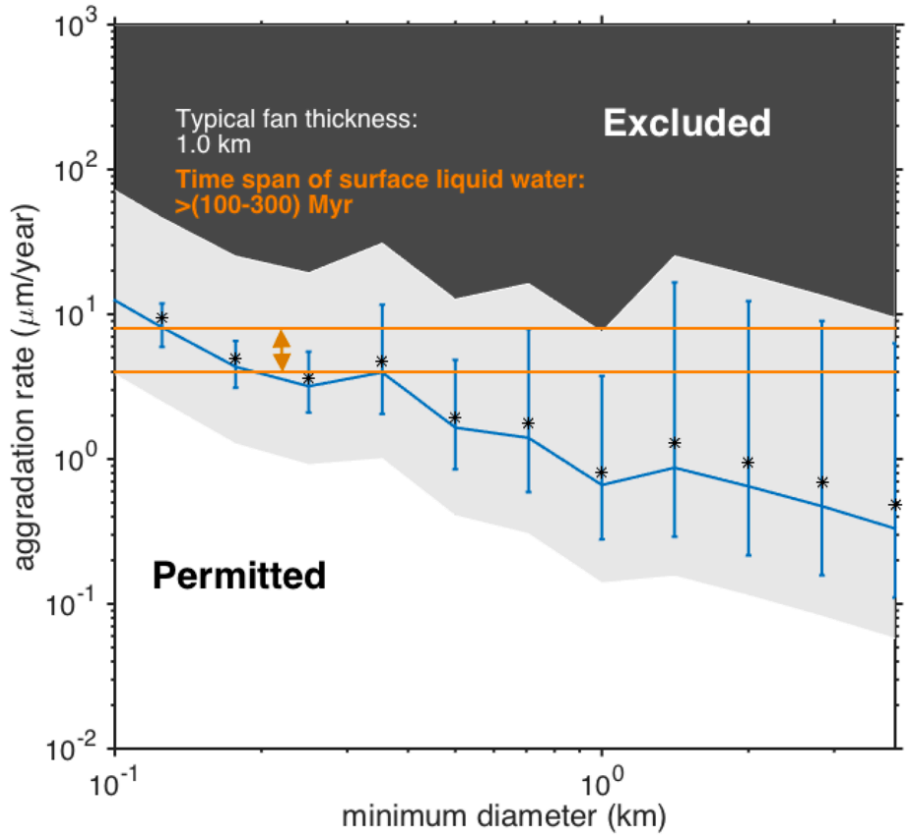

(b)

**Figure 4. (a).** Minimum time-span of sedimentation based on observed synfluvial craters, binned by minimum diameter (cumulative plot). Blue error bars bracket the 90% confidence intervals on lower limit (by Poisson estimation). Full Monte Carlo fit corresponds to the gray band. Black zone is excluded with >95% confidence. White zone is excluded in <5% of trials. Black asterisks correspond to the median outcome of the Monte Carlo procedure. Red dashed lines assume a constant aggradation rate and that burial by 10% of a crater's diameter is sufficient to obscure the crater from orbiter-image surveys. **(b)** Corresponding aggradation rate estimates for alluvial fans, binned by minimum diameter (cumulative plot). Error bars are the same as in (a).

Results are plotted excluding craters for which synfluvial versus prefluvial status could not be determined, but including craters of quality score 3. These choices have little effect on our conclusions. That is because the quality-3 embedded craters, and the craters whose synfluvial status is uncertain, correspond to small diameter bins for which our counts are probably incomplete.

## 4. Discussion.

### 4.1. Factors not taken into account would increase our lower limit.

By fitting a single time-span (and separately fitting a single erosion rate) to the fan deposits, we implicitly assume that the probability of finding an embedded crater is spatially uniform. However, this assumption is inconsistent with the clumpiness of the observed spatial distribution of embedded craters. For example, Crater "W" (Kraal et al. 2008) has 20% of the embedded craters (5 out of 25) even though the fan area at that site is only 500 km$^2$ (3% of total). This cluster is unlikely to be due to chance (assuming cratering is a Poisson process): if fan area is divided into (17000 km$^2$)/(500 km$^2$) = 34 equal-area sites, the expected number of craters per site is $\lambda$=25/34=0.73. The probability of finding ≥5 embedded craters in ≥1 sites is then only $1-(1-\Sigma_{x=5}^{\infty}f(x|0.73))^{34}$ =3%, where $f$ is the Poisson probability distribution function. To be consistent with data at the 50% level, we must increase $\lambda$ by 200%. This might correspond to uniform fan age with spatially non-uniform detectability of embedded craters. Alternatively Crater "W" might record anomalously slow aggradation. In either case, our best estimate of the time span of fan-forming climates is 200% longer than in our lower limit.

Spatial staggering of fan aggradation would increase our lower limit. Time-varying orbital forcing would favor snowmelt (e.g. Kite et al. 2013b) in different places at different times. Localized precipitation would not be globally correlated (Kite et al. 2011).

If the fan deposits underwent erosion during the period of net aggradation, then this would destroy some embedded craters. Therefore, if erosion occurred during fan aggradation, our upper limit would increase further.

Mars crater fluxes are extrapolated from Lunar radiogenic ages and corresponding crater densities (Neukum et al. 2001). Those crater densities have been argued to be incorrect (Robbins 2014). Adopting Robbins' chronology would reduce flux uncertainty from (1-12)× modern (Neukum chronology function) to (1.5 – 2.2)× modern. Because high aggradation rates in our Monte Carlo runs always correspond to high crater fluxes, including Robbins' chronology would raise our lower limit.

## 4.2. Implications for paleohydrology and climate.

$D$<100m embedded craters place an upper limit on paleoatmospheric pressure (Kite et al. 2014). Since $D$<50m impact craters are extremely rare on Earth (atmospheric column density $10^4$ kg/m²), the existence of $D$<100m embedded craters in the Mars fans suggests atmospheric column density <2×$10^4$ kg/m², i.e. $P$<1 bar around the time of fan aggradation (Fig. 5).

Previous analyses of the interval over which alluvial fans formed have divided fan volume by the inferred fluvial sediment transport flux (e.g. Jerolmack et al. 2004). This duration is a lower bound on the interval over which alluvial fans formed, because not all years need produce runoff. Our lower bound exceeds sedimentological lower bounds by >1000-fold. Many alluvial fans are ∼1km thick. Suppose a fan:alcove area ratio of 0.5. Typical water:sediment ratios on Earth are $10^3$:1. 2000km of water at 0.5m snowmelt/year gives 4 Myr. These calculations are highly uncertain. For example, if the amount of snowmelt is limited by snow supply to 10 cm/yr, then the fan formation time is 20 Myr, equal to our strict lower limit on the total time span of alluvial fan formation. However, given quasi-periodic orbital variability, fine-tuning of Mars' hydrological cycle is required to produce small amounts of runoff every year, especially for our preferred lower limit of >(100-300) Myr. Slow net aggradation rates in areas of steep relief (Fig. 1) suggest intermittency.

Intermittency in alluvial-fan-forming climate is further suggested by combining our data with other constraints. The paucity of mineralogic evidence for in-situ alteration of fan deposits (McKeown et al. 2013), the presence of hydrated silica (possibly opal; Carter et al. 2012), and the persistence of olivine (Stopar et al. 2006), when combined with the >20 Myr span of surface liquid water required by our data, suggest that climate conditions were cold and that intermittency further reduced liquid water interaction with soil. Cold conditions are also suggested by sedimentary-deposit mineralogy at Gale (McLennan et al. 2014, Siebach et al. 2017). Intermittency is also suggested by multiple pulses of fan formation at Holden (Irwin et al. 2008), Gale, and Melas Chasma (Williams & Weitz 2014).

In summary, the data exclude any explanation that produces a single burst of habitability of <20 Myr duration. For example, the data exclude triggering by the thermal pulse caused by the impacts that formed the large craters which host the alluvial fans. The data disfavor fluvial sediment transport every year for >20 Myr. Among other possibilities, the data permit a long-lived habitable environment (snowmelt or rainfall); a chaos trigger (Baker et al. 1991); or obliquity-paced fluvial intermittency (Kite et al. 2013b).

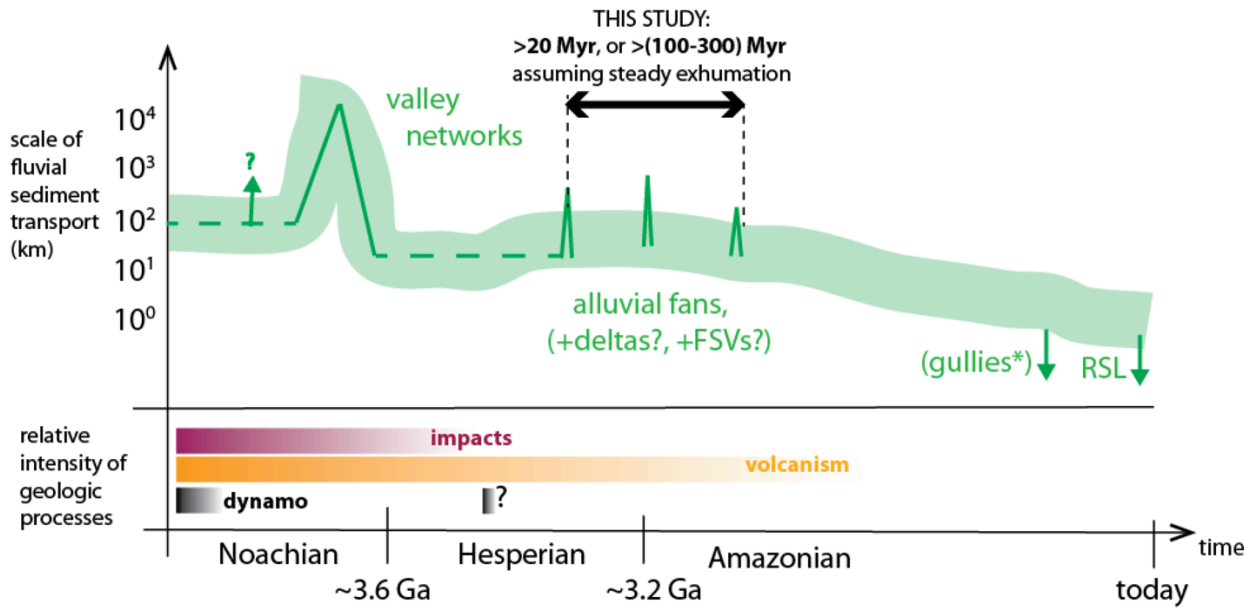

**Figure 5.** Geomorphic history of Mars (Howard 2007, Fassett & Head 2011). FSV = Fresh Shallow Valleys (Wilson et al. 2016). RSL = Recurring Slope Lineae. Pre-valley network fluvial sediment transport is from Irwin et al. (2013).

### 4.3. Implications for habitability.

The time span of alluvial-fan aggradation (§3) is a proxy for the time span of spatially associated paleolakes. Those paleolakes include candidate playa deposits at the toes of fans (e.g. Morgan et al. 2014), a >100m deep paleolake in Crater "P" (Kraal et al. 2008) suggested by a common fan-frontal-scarp elevation of -2700m (in our CTX DTMs), and the Eberswalde paleolake (which shares a drainage divide with fans at Holden; Irwin et al. 2015). In addition, rivers and lakes occurred during the Late Hesperian-Amazonian in Valles Marineris (e.g. Mangold et al. 2004) and Arabia Terra (Wilson et al. 2016). Lake deposits have good biosignature recovery potential (Summons et al. 2011), and biosignature recovery from Proterozoic lake deposits is routine (Peters et al. 2005). However, biosignatures would have been destroyed if lake waters were oxidizing.

### 5. Conclusions.

Explaining young alluvial fans on Mars is a challenge to climate models. To determine the time span of alluvial fan forming conditions, we counted embedded craters. We found a high density of embedded craters, which requires that the river-permitting climate(s) spanned >20 Myr. If aggradation was steady at <(4-8) µm/yr, which is consistent with our data, then fan build-up required >(100-300) Myr (Table 1). The data make the challenge of explaining the alluvial fans more severe, because they exclude a single short-lived anomaly as the cause of the alluvial fans.

**Supplementary Information.** The Supplementary Information is too large to be included in the arXiv version of our article, but may be freely downloaded from the publisher's website via doi:10.1002/2017GL072660 , or from the website of the lead author. This supporting section provides an overview of the prior probabilities used in the Monte Carlo estimation as Text S1. Figure S1 provides a global map for each interbedded crater used in this study. Figure S2 provides images, location information, and ratings for each interbedded crater considered during the CTX-based search. Figure S3 provides comparable information regarding the supplementary HiRISE-based interbedded crater search.

## Acknowledgements.

We thank Alan Howard, Marisa Palucis, Becky Williams, Ross Irwin, Jean-Pierre Williams, Bill Dietrich, and Brian Hynek. We thank two anonymous reviewers for their useful recommendations, and Andrew Dombard for editorial handling. All data may be obtained by contacting the lead author (kite@uchicago.edu). We are grateful for financial support from the US taxpayer via NASA grants NNX16AG55G and NNX15AM49G.

## References.